\documentclass{PoS}

\title{Characterization of a Maximum Likelihood Gamma-Ray Reconstruction Algorithm for VERITAS}

\ShortTitle{VERITAS Reconstruction Algorithm}

\author{\speaker{Jodi Christiansen}, for the VERITAS Collaboration\thanks{veritas.sao.arizona.edu} \\
	California Polytechnic State University, San Luis Obispo\\
	E-mail: \email{jlchrist@calpoly.edu}}

\abstract{We characterize the improved angular and energy resolution of a new likelihood gamma-ray reconstruction for VERITAS.  The algorithm uses the average photoelectrons stored in templates that are based on simulations of large numbers of showers as a function of 5 gamma-ray parameters: energy, zenith angle, core location (x,y), and depth of first interaction in the atmosphere.  Comparing the template predictions of the average photoelectrons in each pixel to observed photoelectrons allows us to calculate the likelihood.  By maximizing the likelihood, we find the optimal gamma-ray parameters.  The maximum likelihood reconstruction improves on the standard VERITAS analysis which relies on: 1. the weighted average of the axis of elongation in the images to determine the gamma-ray direction and 2. look-up tables that relate the observed energy deposition of Cherenkov photons to the true gamma-ray energy.  Not only is the maximum likelihood method more accurate, but it is also not biased by missing pixel information due to the edge of the camera or pixel cleaning. The drawback is that it takes more CPU time (80 ms/event).}

\FullConference{35th International Cosmic Ray Conference --- ICRC2017\\
		10--20 July, 2017\\
		Bexco, Busan, Korea}

\begin{document}

\section{Introduction}

Imaging Air Cherenkov Telescopes (IACTs) have been using weighted averages of air-shower images to successfully reconstruct gamma-rays for over twenty years.  Hillas parameters \cite{Hillas} are formed from the moments of the shower-image formed by Cherenkov light from air-showers.  The direction of these moments is traditionally averaged to find the air-shower direction, and look-up tables are typically used to relate the amount of the detected light to the incident $\gamma$-ray energy \cite{VERITASHillas}.  As processor speed and memory increase, maximum likelihood fitting methods have become a viable option.  LeBohec and de Naurois have demonstrated the improved resolution and sensitivity of maximum likelihood methods using shower-image templates for IACTs \cite{LeBohec, Naurois} and the H.E.S.S. collaboration has created the ImPACT algorithm based on the method \cite{Parsons}.  Here we report on our implementation of the shower-image template maximum likelihood $\gamma$-ray reconstruction method (ITM) in a VERITAS standard analysis package (VEGAS\_v2.5.6)\cite{VEGAS}.  The primary advantage of maximum likelihood methods is that they are not biased by missing information such as images that extend beyond the edge of the camera or pixels removed because of bright starlight.
\setcounter{figure}{0}
\begin{figure}[b]
	\includegraphics[width=150mm]{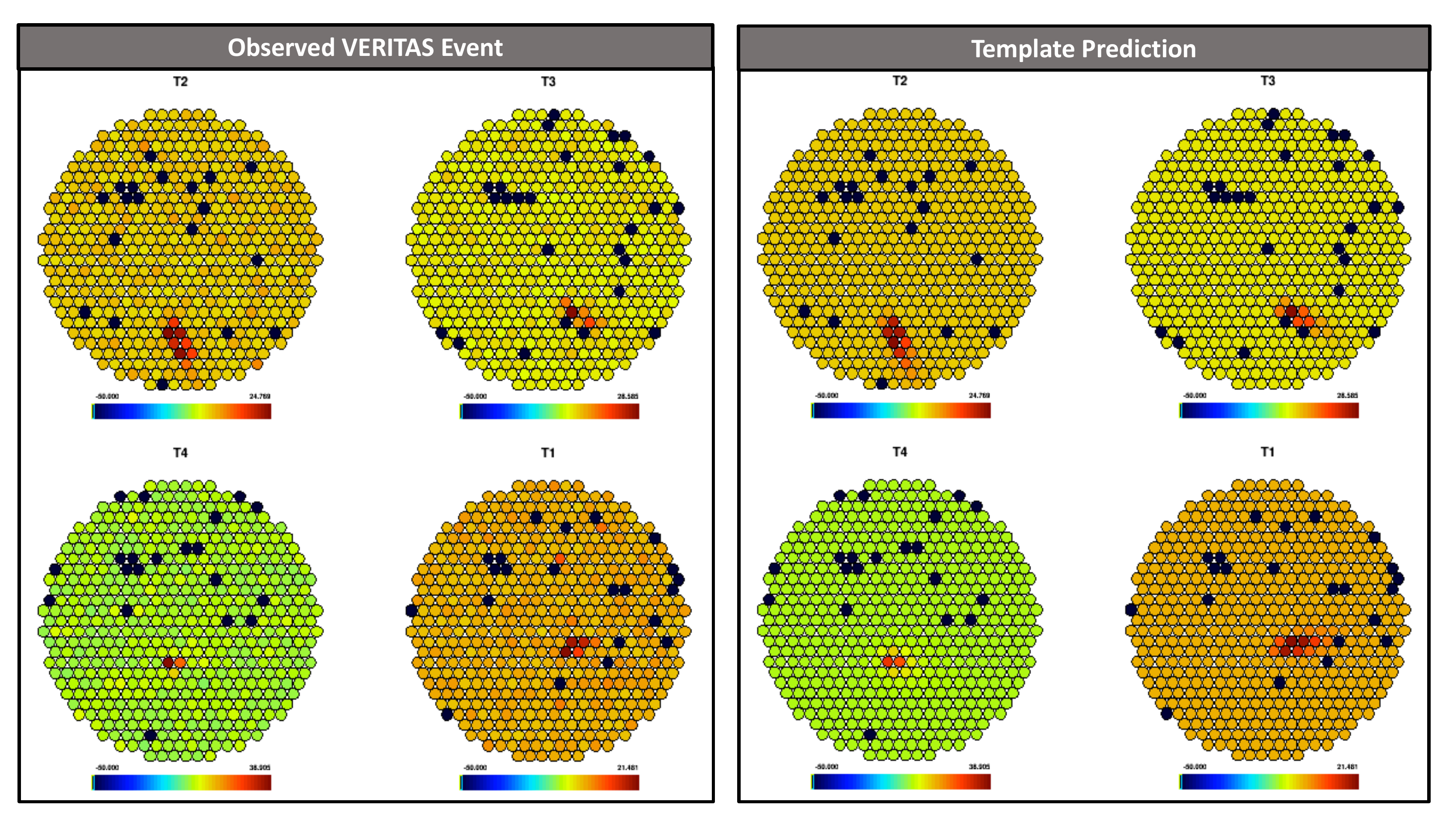}
	\caption{Images from each of the four VERITAS telescopes. (left) Real candidate $\gamma$-ray event, (right) Image-template prediction.}
	\label{fig:Event}
\end{figure}

Our method interpolates between shower-image templates based on six gamma-ray parameters: energy, zenith angle, azimuthal angle, core location (x,y), and depth of first interaction in the atmosphere.  The templates were created by Vincent \cite{Vincent} for VERITAS using CORSIKA simulations of gamma-ray air showers .  Each template contains the average number of photoelectrons that are expected to be detected by a VERITAS PMT based on the incident energy, zenith angle, core location, as well as the depth of first interaction in the atmosphere. The reconstruction code rotates the template pixels to the camera coordinates for a specific azimuth and wobble.   It then calculates the likelihood by comparing the template prediction to the observed number of photoelectrons in each pixel. By maximizing the likelihood, we find the optimal $\gamma$-ray parameters. Example  camera images are shown in Fig.~\ref{fig:Event}.  Here, a low-energy $\gamma$-ray candidate event is shown on the left and the template prediction is shown on the right.  The reconstructed energy for this event is about 150 GeV, close to the energy threshold for VERITAS.  Missing pixels are shown in black.  The group of four pixels in the NW portion of each camera are coincident with the image of a bright star. To reduce damaging anode currents in the photomultiplier tube, the HV in these pixels was lowered during the observation.  It is not unusual for the missing pixels to affect the $\gamma$-ray reconstruction.

\setcounter{figure}{1}
\begin{figure}[b]
	\includegraphics[width=150mm]{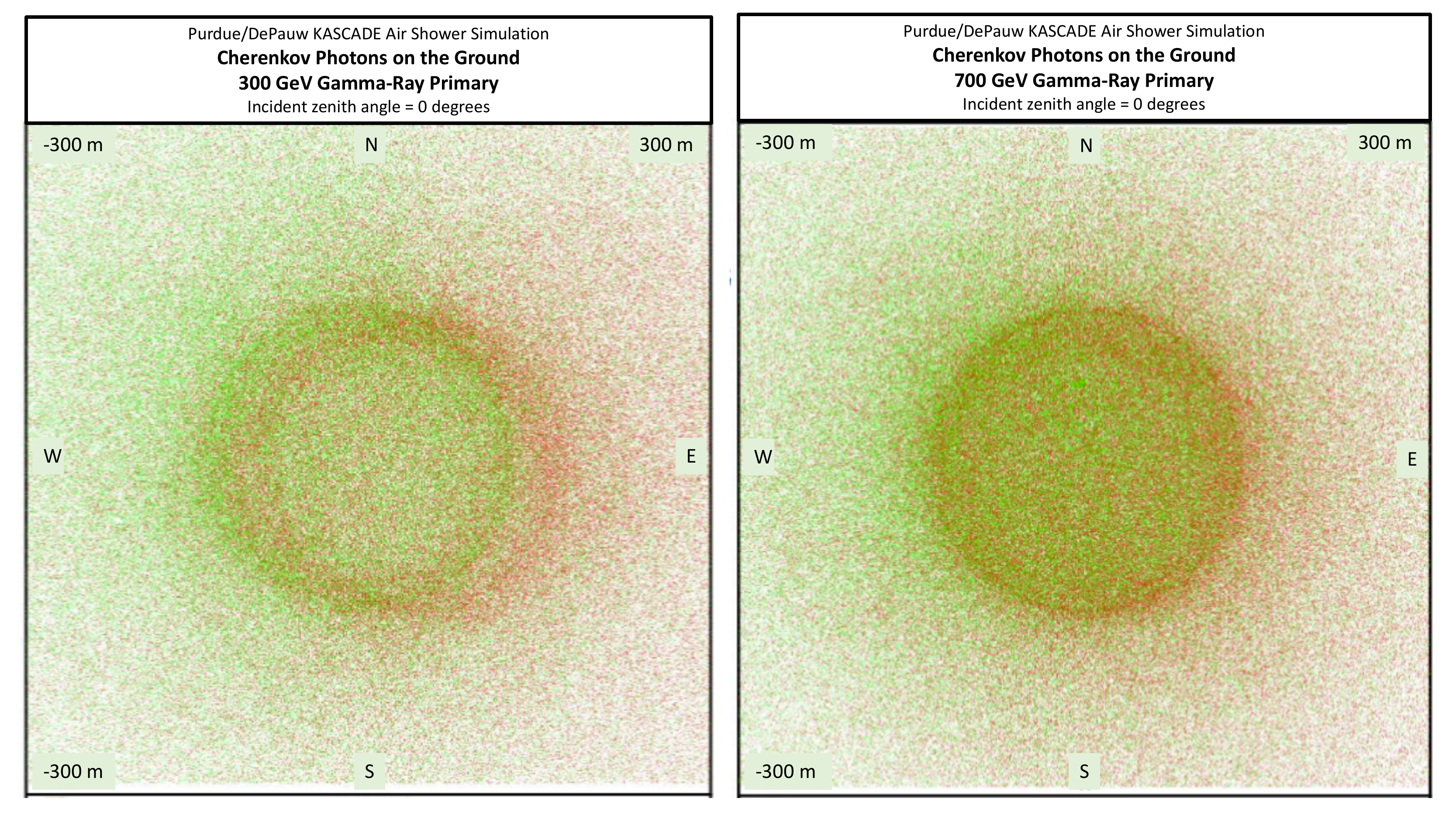}
	\caption{Simulated Cherenkov photons reaching the ground from vertical air-showers by 300 GeV (left) and 700 GeV (right) $\gamma$-ray primaries \protect\cite{Kascade}.  Photons radiated by electrons are shown in green.  Photons radiated by positrons are shown in red. Their separation is due to the Lorentz force generated by Earth's magnetic field.}
	\label{fig:MagField}
\end{figure}

To make fitting methods robust and fast, we consider four practicalities of the implementation.  First, maximum likelihood methods require that the likelihood produces a reliable gradient so that they move successfully to the optimal solution. For IACTs, the requirement is that the seed parameters produce a pool of expected Cherenkov light that overlaps with the observed shower image.  Most of the camera is dark, and the gradient is not valid if the seed predicts light only where the pixels contain pedestal noise.  We therefore seed the new algorithm with gamma-ray parameters from the standard moment analysis which are generally close to the optimal parameters.  The second practical consideration is time.  The average time to reconstruct and event is 80 ms.  We use the mean-scaled width and length of the shower to select only the likely gamma-rays for reconstruction.  These cuts, which primarily remove cosmic ray showers, are independent of the reconstructed parameters and do not bias the reconstructed events.  The third practical consideration is the relationship between the template predictions and the observed signals.  An overall energy calibration factor was developed to account for the detector effects, including seasonal differences in the atmosphere, mirror reflectivity and electronic response, among other considerations. Finally, Earth's magnetic field also affects the energy scale, especially for lower-energy showers, which show a separation of electrons and positrons as the shower develops due to the Lorentz force.  As shown in Fig.~\ref{fig:MagField}, by the time they reach the ground, Cherenkov photons radiated by  electrons are separated from those radiated by positrons, changing the density and angular distribution of the photons entering the IACTs.  Because the strength of the force depends on the angle between the incident $\gamma$-ray and the local magnetic field, we apply a calibration correction for the energy that depends on both the incident zenith and azimuthal angles.

\section{Angular and Energy resolution}
The simulated angular resolution is shown in Fig.~\ref{fig:AngRes}.  The  68\% containment radius depends on the $\gamma$-ray energy and zenith angle.  The standard moment analysis is shown for comparison.  The new algorithm localizes $\gamma$-rays 25\% to 30\% better. Angular selection cuts, $\theta$, were optimized for the detection of a 1\% Crab flux using VERITAS data.  For a medium to hard spectrum, with a power-law index less than 3, the optimal cut on angular separation between the reconstructed $\gamma$-ray direction and the direction to the source is $\theta<0.071^{\circ}$.  For soft spectra, with a power-law index greater than 3, the optimal cut is  $\theta<0.1^{\circ}$.  The sensitivity of the new medium to hard analysis is 35\% better than the moment analysis and the soft analysis is 30\% better.
\setcounter{figure}{2}
\begin{figure}[b]
	\includegraphics[width=79mm]{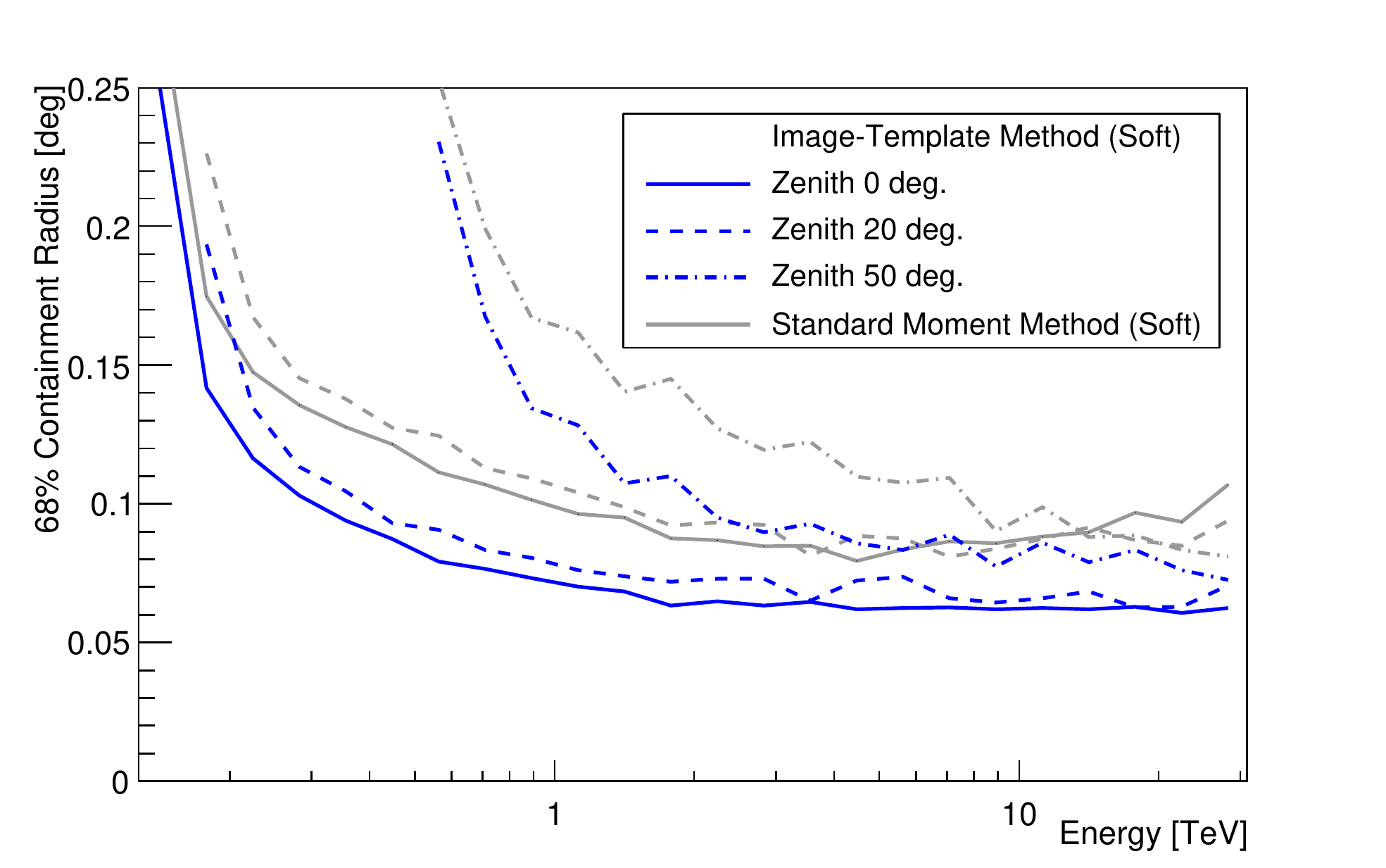}
	\includegraphics[width=79mm]{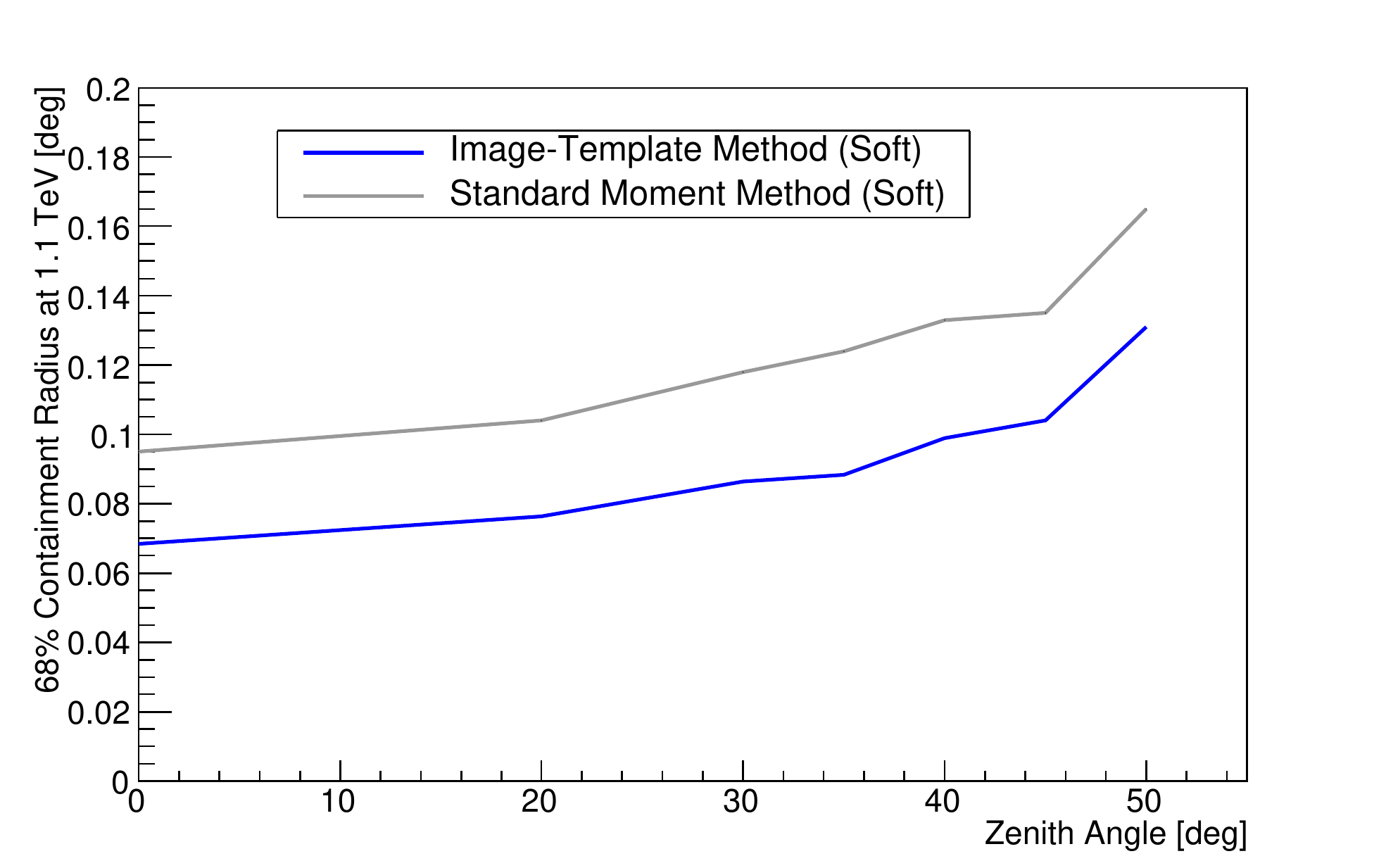}
	\caption{The angular resolution of VERITAS (left) as a function of energy for a fixed zenith angle of 20 degrees and (right) as a function of zenith angle for a fixed energy of 1.1 TeV. The angular resolution of the image-template method is generally 25\% smaller.}
	\label{fig:AngRes}
\end{figure}

\setcounter{figure}{3}
\begin{figure}
	\includegraphics[width=165mm]{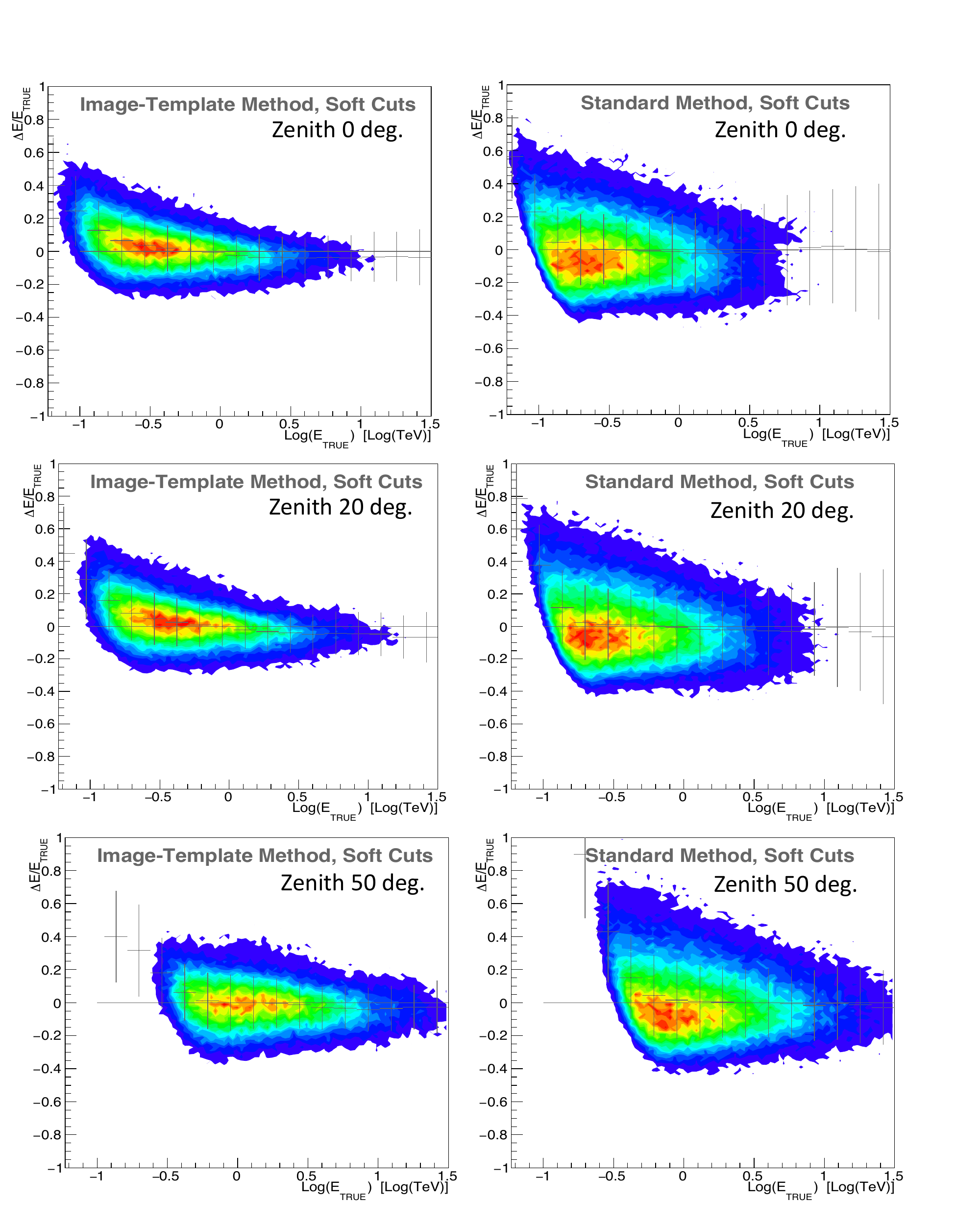}
	\caption{The energy resolution of VERITAS for (left) the image-template method (ITM) and (right) the standard Hillas method. We define $\Delta E = E_{RECONSTRUCTED} - E_{TRUE}$. The standard deviation of the ITM distribution is 24\% smaller.}
	\label{fig:EngRes}
\end{figure}

The simulated energy resolution is shown in Fig.~\ref{fig:EngRes} for a variety of zenith angles.  At the lowest energies, the simulated trigger bias is evident.  At energies above a few TeV, some pixels are likely to switch to their low-gain mode to avoid saturating.  The templates have not yet been tested for pixels in the low-gain mode with $N_{pe}>300$, so at present the new algorithm removes these pixels from the reconstruction.  The energy can still be reconstructed with minimal bias using the remaining pixels in the image.  The energy range of the new algorithm extends from 140 GeV to 30 TeV.  The trigger bias is currently folded into the systematic uncertainty, but we have plans to implement an unfolding algorithm for the spectral reconstruction code in the future.

\setcounter{figure}{4}
\begin{figure}[b]
	\includegraphics[width=77mm]{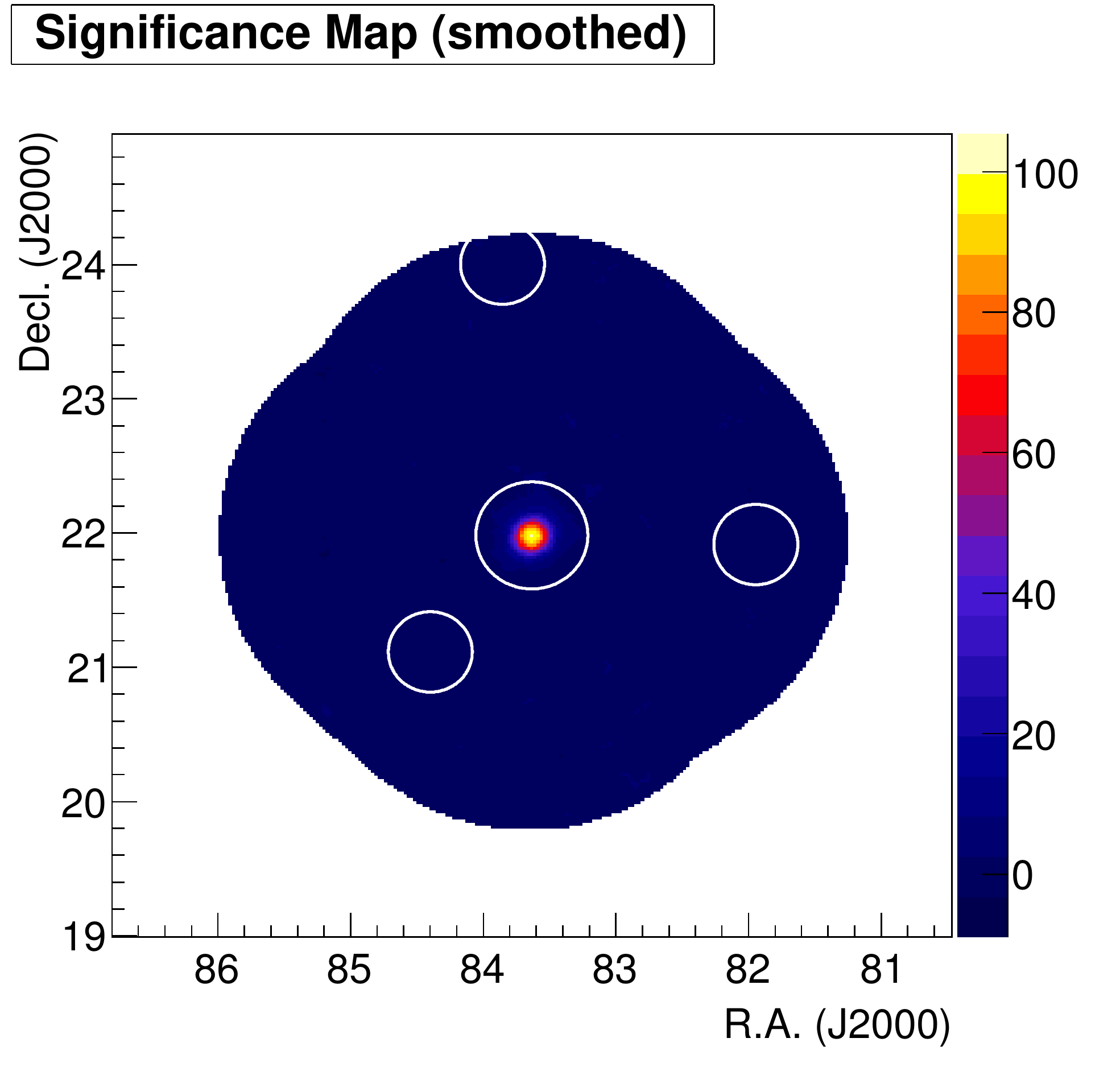}
	\includegraphics[width=77mm]{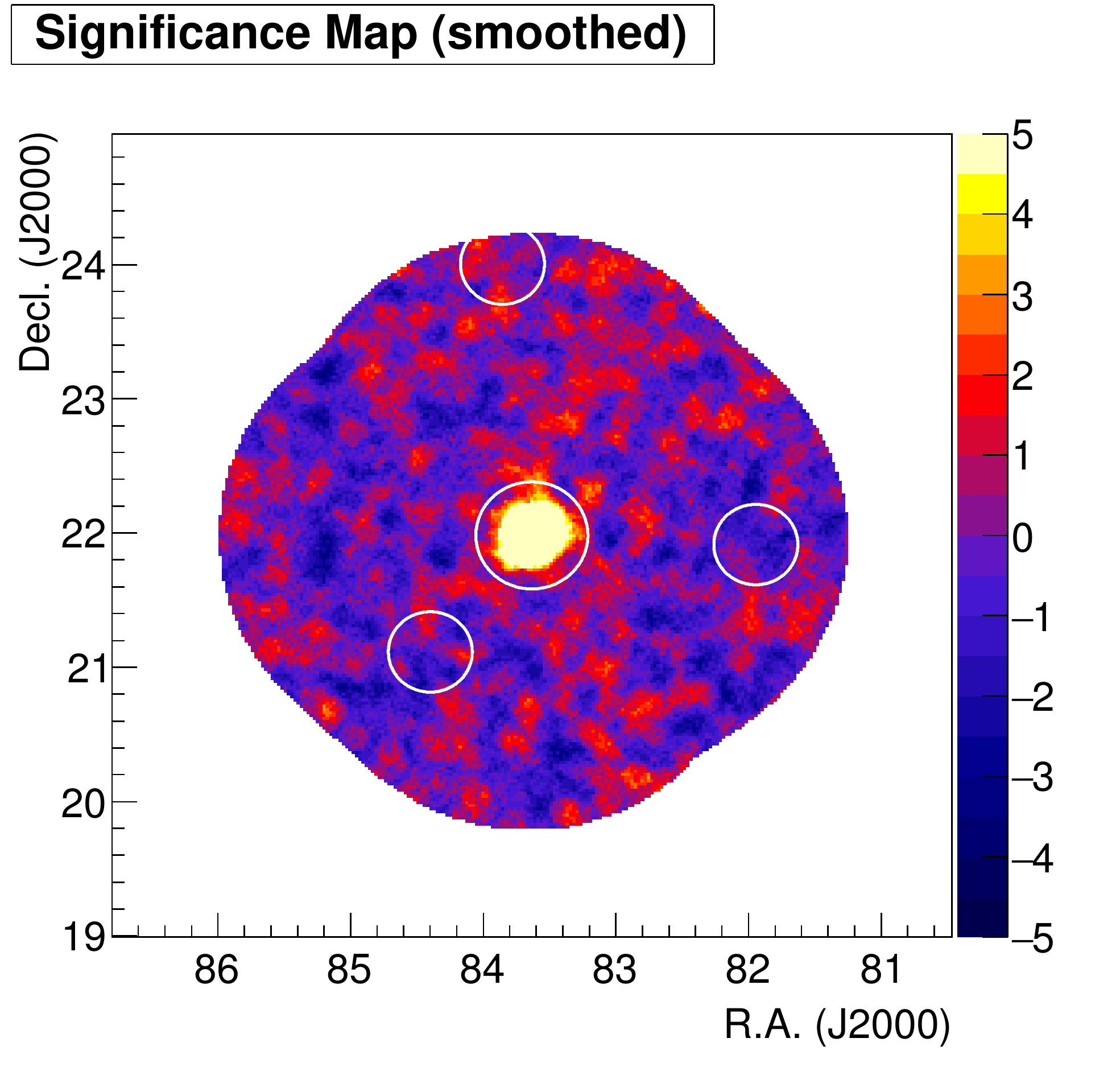}
	\caption{The significance maps for 3.75 hours of Crab observations analyzed with the soft cuts. (left) Source localization is consistent with Fig.~\ref{fig:AngRes}.  (right) Scale set to show background features.  Three stars are in the field. There is no bias evident from these nuisance regions.}
	\label{fig:SigMaps}
\end{figure}

\section{Crab Observations}

We analyzed 3.75 hours of observations taken between 2013 and 2016 to show how the algorithm with soft cuts performs on real data. Significance maps are shown in Fig.~\ref{fig:SigMaps}. The source is well localized and the angular spread of $\gamma$-ray candidates is in good agreement with the simulations.  The significance scale was adjusted on the right-hand map to show that background fluctuations are distributed as expected. Three stars are in the field and there is no evidence of bias in these nuisance regions.  This is confirmed by the significance distribution shown in Fig.~\ref{fig:SigDists}.  Notice that the backgrounds, even those within the circles denoting stars follow a Gaussian function. The analysis of the Crab spectrum shows good agreement with the standard algorithm.

\setcounter{figure}{5}
\begin{figure}[t]
	\centering
	\includegraphics[width=75mm]{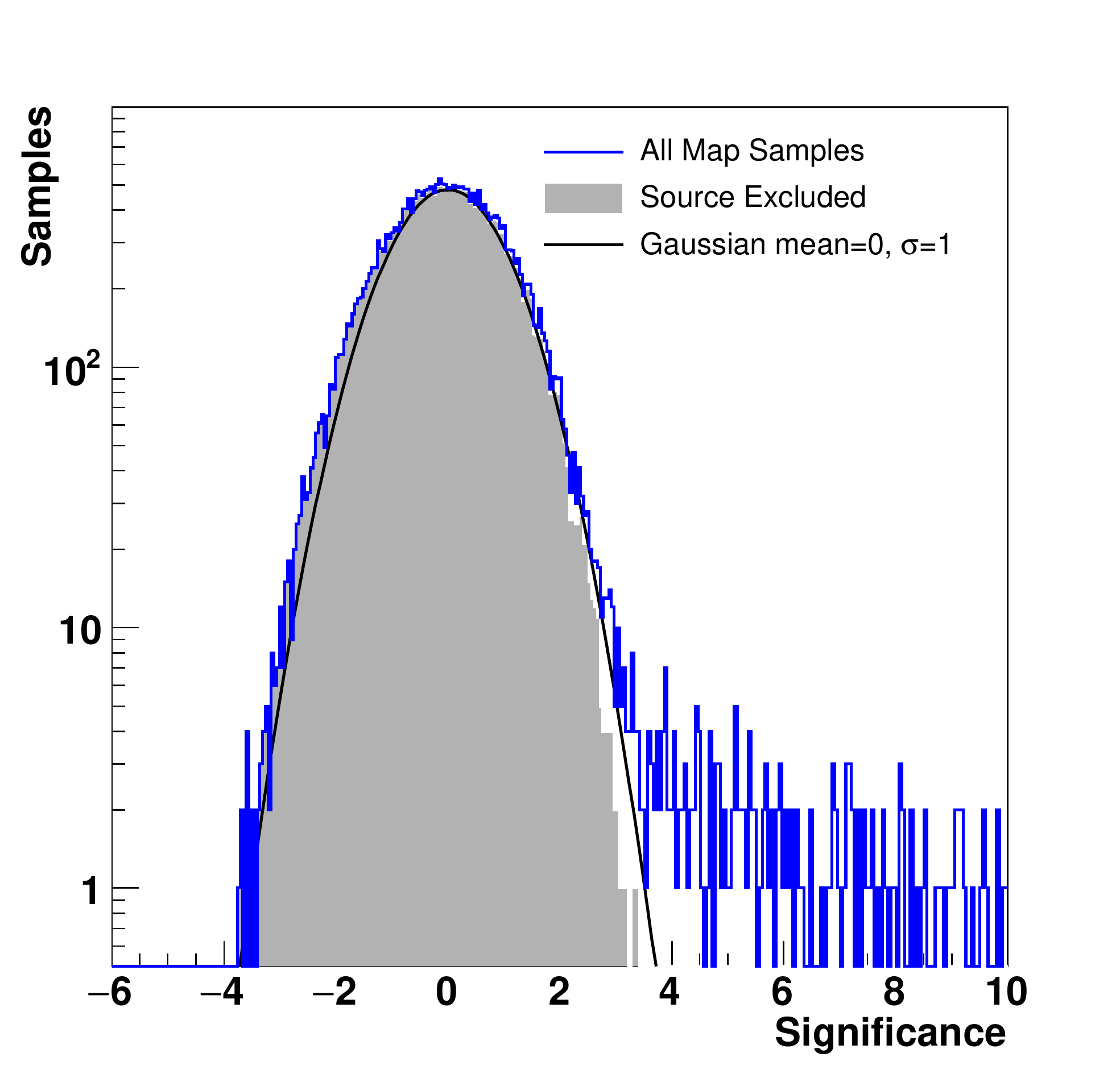}
	\caption{(The significance distribution shows that the backgrounds are consistent with a Gaussian model centered at zero with a width of one. The background samples include the regions with stars.}
	\label{fig:SigDists}
\end{figure}
\setcounter{figure}{6}
\begin{figure}[b]
	\centering
	\includegraphics[width=75mm]{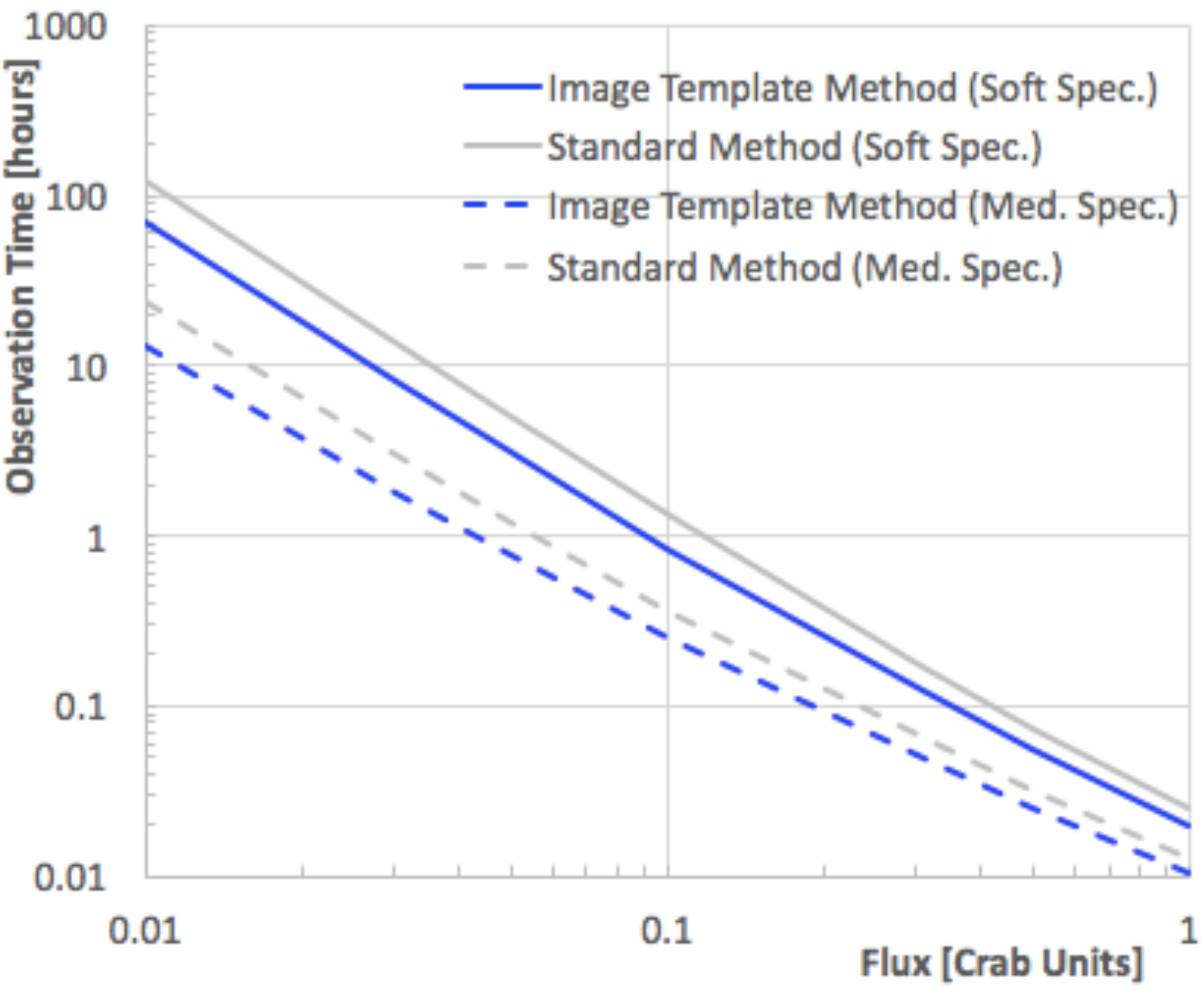}
	\caption{Observation time to detect a point-source at 5$\sigma$ significance as a function of flux in Crab Units. We define the Crab Unit as the integral flux of the Crab Nebula above 180 GeV. A Crab Unit = $2.46\times10^6$ photons m$^{-1}$ s$^{-1}$ using the Crab Nebula flux reported by the MAGIC Collaboration \protect\cite{Magic}. Dashed lines were determined by scaling Crab Nebula observations. Solid lines correspond to scaling PKS 1424+240 observaions. }
	\label{fig:ObsTime}
\end{figure}

\section{Sensitivity}
The optimal angular cuts are more stringent now that the angular resolution is improved.  The tighter angular cuts result in higher sensitivity primarily due to reduced background from a smaller on-source region. The smaller nuisance regions near stars also provide more area to measure the backgrounds accurately.  Fig.~\ref{fig:ObsTime} shows the time it takes to detect a source for various fluxes.  For simplicity, the flux is given in Crab Units (CU) defined to be the integral flux of the Crab Nebula above 180 GeV.  The medium spectrum used here is scaled using Crab observations (spectral index of 2.48) \cite{Magic} and the soft spectrum is scaled using PKS 1424+240 observations (spectral index of 3.8, energy threshold 140 GeV) \cite{PKS}. With the image-template method, dim sources with an intensity of 1\% CU can be detected more than 40\% quicker. These improvements have come entirely through the application of more sophisticated software and are comparable to the improvements achieved by the camera upgrade in 2012 \cite{Nahee}.

\section{Conclusions}
The  new image-template method provides a significant improvement to the science reach of VERITAS. The sensitivity is improved by 25\% for dim soft-spectrum sources and 35\% for dim medium- to hard-spectrum sources.  The background nuisance regions around stars are substantially reduced making analysis far simpler. The improved energy resolution is expected to reduce uncertainties in spectral parameters. Furthermore, the algorithm is also practical to run, requiring only about 2.5 times more disk space at the earliest stages and about 30\% more processing time.

\section{Acknowledgements}
This research is supported by grants from the U.S. Department of Energy Office of Science, the U.S. National Science Foundation and the Smithsonian Institution, and by NSERC in Canada. This research used computational resources of the National Energy Research Scientific Computing Center, a DOE Office of Science User Facility supported by the Office of Science of the U.S. Department of Energy under Contract No. DE-AC02-05CH11231. We acknowledge the excellent work of the technical support staff at the Fred Lawrence Whipple Observatory and at the collaborating institutions in the construction and operation of the instrument. The VERITAS Collaboration is grateful to Trevor Weekes for his seminal contributions and leadership in the field of VHE gamma-ray astrophysics, which made this study possible.

\end{document}